\documentclass[11pt,journal,compsoc]{IEEEtran}

\ifCLASSOPTIONcompsoc

\else

\fi

\usepackage{mathrsfs}
\usepackage{amsmath}
\usepackage{amssymb}
\usepackage{amsthm}

\ifCLASSINFOpdf

\else

\fi

\usepackage[dvips]{graphicx}
\usepackage{algorithm}
\usepackage{algorithmic}

\begin{document}

\title{A Raindrop Algorithm for Searching The Global Optimal Solution in Non-linear Programming}

 \author{
 \IEEEauthorblockN{Zhiqing Wei\\}
 \IEEEauthorblockA{Beijing Univ. of Posts and Telecom. (BUPT)\\
 Beijing, P. R. China, 100876\\
 email: zhiqingwei@gmail.com}}
\maketitle

\begin{abstract}
In this paper, we apply the random walk model in designing a raindrop algorithm to find the global optimal solution of a non-linear programming problem. The raindrop algorithm does not require the information of the first or second order derivatives of the object function. Hence it is a direct method. We investigate the properties of raindrop algorithm. Besides, we apply the raindrop algorithm to solve a non-linear optimization problem, where the object function is highly irregular (neither convex nor concave). And the global optimal solution can be found with small number of iterations.
\end{abstract}

\IEEEpeerreviewmaketitle

\section{Introduction}

The method to find the global optimal solution of a general 
non-linear programming does not exist \cite{Convex_optimization}. 
Thus some randomized and heuristic algorithms \cite{Alg_Design} 
are proposed to find the local optimal or near global optimal solution. 
In \cite{Alg_Design} (Chap. 13), various randomized algorithms 
are listed to solve the combinatorial optimization and function 
optimization problems. The randomized algorithms can 
find the global optimal solution with a probability $p_s$, 
which is defined as success probability to find the global optimal solution. 
And generally, $P_s$ will tend to $1$ when the time 
or space complexity tends to infinity.

In the family of randomized algorithms, the swarm intelligence algorithms \cite{Swarm_alg} are famous for their efficiency in solving the combinatorial optimization and function optimization problems. Among them, the ant colony optimization exploits the multi-agents method and find the optimal solution by locally exchange the information of solution among the artificial ants. Another particle swarm optimization (PSO) model uses the local optimal and temporary global optimal solution to guide the solution update in each iteration.

In this paper, we solve the non-linear programming with 
a different method and propose a raindrop method, which is inspired 
by the nature of raindrop. When raindrops fall on the ground, 
they will always move to the (local) lowest spots. 
Thus we use a random walk model \cite{Mobility_Model} to control 
the movement of raindrops to find 
the global optimal point (solution), namely,
each raindrop moves with a certain speed on the ground to 
find the optimal solution (lowest point on the ground).

This paper is organized as follows. In section 2, we give the description of the raindrop algorithm. In section 3, some properties of this algorithm are provided. In section 4, we give some numerical results. And the conclusions are in section 5.

\section{Algorithm Description}

\subsection{Random Walk Model}

\begin{figure}[!t]
\centering
\includegraphics[width=0.5\textwidth]{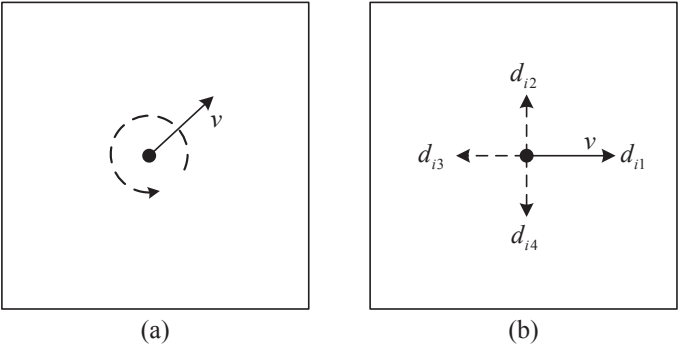}
\caption{Random walk model.}
\label{fig_RW_Model}
\end{figure}

The random walk (RW) model is illustrated in Fig. \ref{fig_RW_Model}. 
In RW model, $N$ nodes are uniformly deployed in a finite region. 
Then at each time interval each node randomly chooses an 
angle $\theta  \sim (0,2\pi ]$ and a velocity $v$, 
and moves in direction $(\cos \theta ,\sin \theta )$ with 
velocity $v$ until the next time interval. 
When the next time interval comes, 
each node updates the direction and velocity following the same rules.
In Fig. \ref{fig_RW_Model} (a), the node can move in continuous direction. 
But in Fig. \ref{fig_RW_Model} (b), the node can only move 
in four directions, which is a simplification of RW model. 
In this paper, the direction of raindrop's movement 
is not random. Suppose we deploy $N$ raindrops on the ``ground'', 
then each raindrop should move in the direction of negative gradient. 
However, we assume the gradient of the object function is not available. 
Thus the raindrop should move in the direction of 
negative quasi gradient direction. 
But we simplify the operation to the greatest extent, 
i.e., the raindrop just needs to move in one of the four directions 
in Fig. \ref{fig_RW_Model} (b) which leads to the smallest 
value of the object function.

\subsection{Raindrop Algorithm}

A general optimization problem is as follow.
\begin{equation}\label{eq_a_opti_problem}
\begin{aligned}
& \mathop {\min }\limits_{x \in {R^n}} f(x) \\
& s.t.{\kern 3pt} x \in S \subset {R^n}
\end{aligned}
\end{equation}
where x is an n-dimensional vector and $S$ is a set with finite measure. We design a raindrop algorithm to solve this general optimization problem.

Firstly, $N$ raindrops randomly fall on the ``ground'', where the ``ground''
is $S$ in (\ref{eq_a_opti_problem}). We denote the location of $i$th
raindrop as $x_i \in S$. After the raindrop falls,
they will move at each time interval.
In time interval $j$, we suppose the raindrop can
moves in one of the $2n$ directions as follows.

\begin{equation}
{d_{i,k}} = \left\{ {\begin{array}{*{20}{c}}
   {{v_{i}}{e_k}} & {k \le n}  \\
   { - {v_{i}}{e_{k - n}}} & {n < k \le 2n}  \\
\end{array}} \right.
\end{equation}
where $[{e_1},{e_2}, \cdots ,{e_n}] = I_n$ 
with $I_n$ an $n \times n$ identity matrix. 
The $d_{i,k}$'s are illustrated in 
Fig. \ref{fig_RW_Model} (b) for $n = 2$. $v_i$ is 
the velocity of raindrop $i$.

Each raindrop moves in one of the $2n$ directions which leads to 
the smallest value of the object function, i.e., 
the next (candidate) movement direction of the raindrop in the next time interval is

\begin{equation}\label{eq_direction}
{d_{i,j + 1}} = \mathop {\arg \min }\limits_{{d_{i,k}}} f({x_{i,j}} + {v_{i,j}}{d_{i,k}})
\end{equation}

If $f({x_{i,j}} + {v_i}{d_{i,j + 1}}) < f({x_{i,j}})$, then the location of $i$th raindrop in the next time interval should be

\begin{equation}\label{eq_location_update}
{x_{i,j + 1}} = {x_{i,j}} + {v_i}{d_{i,j + 1}}
\end{equation}
Otherwise, namely, 
when $f({x_{i,j}} + {v_i}{d_{i,j + 1}}) \ge f({x_{i,j}})$, 
which means $x_{i,j}$ is near a local optimal solution, 
then this raindrop should slow down to find 
the accurate (local) optimal solution, i.e., 
the velocity of $i$th raindrop is modified as

\begin{equation}\label{eq_velocity_update}
{v_i}: = \frac{{{v_i}}}{2}
\end{equation}

When the velocity of all the raindrop tends to $0$, namely, 
all the raindrops stop moving, 
then we find all the local optimal solutions, 
which contain the global optimal solution. 
The raindrop algorithm is described in Algorithm 1.

\algsetup{indent=2em}
\begin{algorithm}
\caption{Raindrop Algorithm}

\label{alg3}
\begin{algorithmic}[1]

\STATE $N$ raindrops randomly fall on $S$. The location and velocity of $i$th raindrop are $x_i \in S$ and $v_i$ respectively. Initialize a $0 < \varepsilon  \ll 1$. Assume $x = [{x_1},{x_2}, \cdots ,{x_N}]$ and $v = [{v_1},{v_2}, \cdots ,{v_N}]$.

\WHILE{${\left\| v \right\|_2} > \varepsilon$}

\FOR{$i$th raindrop}

\STATE Find the candidate direction by (\ref{eq_direction}).

\IF{$f({x_{i,j}} + {v_i}{d_{i,j + 1}}) < f({x_{i,j}})$}

\STATE Update location by (\ref{eq_location_update}).

\ELSE

\STATE Update the velocity by (\ref{eq_velocity_update}).

\ENDIF

\ENDFOR

\ENDWHILE

\STATE Find the global optimal solution among the local optimal solutions.

\end{algorithmic}
\end{algorithm}

\section{Algorithm Analysis}

In this section, we will analyze the time and space complexity 
of the raindrop algorithm and show the properties of success probability.

\subsection{The time and space complexity}

The space complexity is $\Theta (N)$, 
which is the same order as the number of raindrops. 
The time complexity is similar to the method of bisection, 
which is $\Theta (N\log \frac{v}{\varepsilon })$. 
Thus with the decrease of $\varepsilon$, namely, 
the accuracy requirement is high, the time complexity will increase.

\subsection{Success probability}

The term of \emph{search vicinity} is defined as a region, 
where if a raindrop falls in, this raindrop can move to the global 
optimal point. The search vicinity is around the 
global optimal solution ${x^*}$ and contains only one global 
optimal solution (contains no other local optimal solutions). 
And the \emph{maximum search vicinity} is defined as the region where if 
a raindrop does not fall in, then this raindrop will not 
moves to the global optimal point (solution).

Therefore, if the distances between the global optimal 
solution and other local optimal solutions do not tend to 0, 
then the measure of the maximum search vicinity is not $0$. 
Because we define the measure of $S$ as finite,
the probability that a raindrop falls in the 
maximum search vicinity $T$ 
is $p = \frac{{\left| T \right|}}{{\left| S \right|}}$, 
where $\left| * \right|$ denotes the measure of set $*$.

When no raindrops fall in the maximum search vicinity $T$,
the raindrop algorithm cannot find the global optimal solution. 
The probability of this event 
is ${p_0} = {\left( {1 - \frac{{\left| T \right|}}{{\left| S \right|}}} \right)^N}$. 
Thus the success probability, namely, 
the probability that the raindrop algorithm can 
find the global optimal solution, is

\begin{equation}
{p_s} = 1 - {\left( {1 - \frac{{\left| T \right|}}{{\left| S \right|}}} \right)^N}
\end{equation}

Obviously, $\mathop {\lim }\limits_{N \to \infty } {p_s} = 1$. 
Actually, in application, a small $N$ can also ensure a considerable $p_s$.

\section{Numerical Results}

\begin{figure}[!t]
\centering
\includegraphics[width=0.5\textwidth]{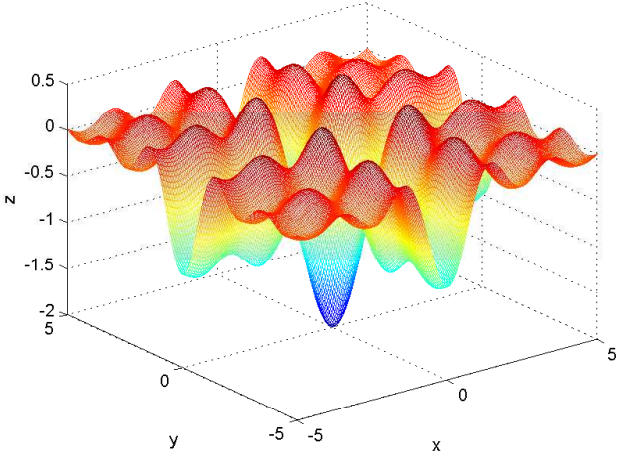}
\caption{The object function of (\ref{eq_example_problem_sinc}).}
\label{fig_obj}
\end{figure}

We use the raindrop algorithm to solve an optimization problem as follow.

\begin{equation}\label{eq_example_problem_sinc}
\begin{aligned}
& \min  {\kern 3pt}  - \frac{{\sin x}}{x} - \frac{{\sin y}}{y} \\
& s.t.{\kern 4pt}  - 5 \le x \le 5, - 5 \le y \le 5 \\
\end{aligned}
\end{equation}

\begin{figure}[!t]
\centering
\includegraphics[width=0.5\textwidth]{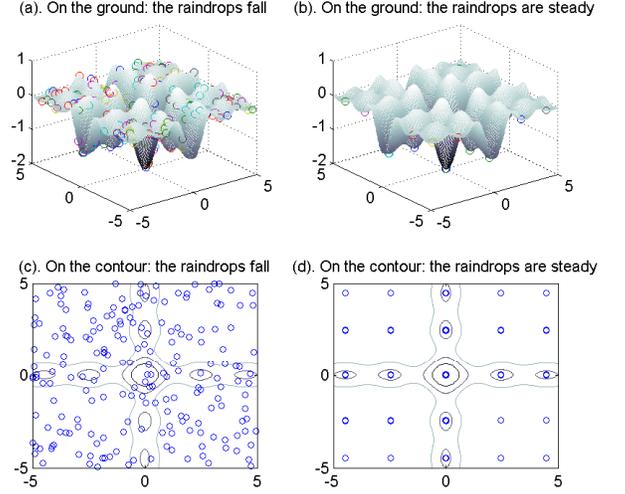}
\caption{Optimization results: (a) and (b) illustrate the procedure that the raindrops falls on the ``ground'' and move to a steady state, where the velocity of raindrops tend to 0. (c) and (d) illustrate the corresponding situations of (a) and (b) on the contour.}
\label{fig_rain_fall_steady}
\end{figure}

The object function is illustrated in Fig. \ref{fig_obj}, 
the optimal solution is $(x^*,y^*) = (0,0)$. 
We apply the raindrop algorithm to solve (\ref{eq_example_problem_sinc}). 
In Fig. \ref{fig_rain_fall_steady}, we plot the procedure of the algorithm. 
In Fig. \ref{fig_rain_fall_steady} (a), 
the raindrops randomly fall on the ``ground'', i.e., 
the set $S$. Then we run the raindrop algorithm, 
the raindrops are moving to find the lowest place to stay. 
When the sum of the velocity of all raindrops tends to $0$, 
the raindrops are steady and the local optimal solutions are obtained. 
We find the global optimal solution among them. 
In Fig. \ref{fig_rain_fall_steady} (b), the raindrops are steady, 
namely, the velocity tends to $0$, then the algorithm can stop. 
Fig. \ref{fig_rain_fall_steady} (c) and (d) 
illustrate the corresponding situations of (a) and (b) on the contour.

\begin{figure}[!t]
\centering
\includegraphics[width=0.5\textwidth]{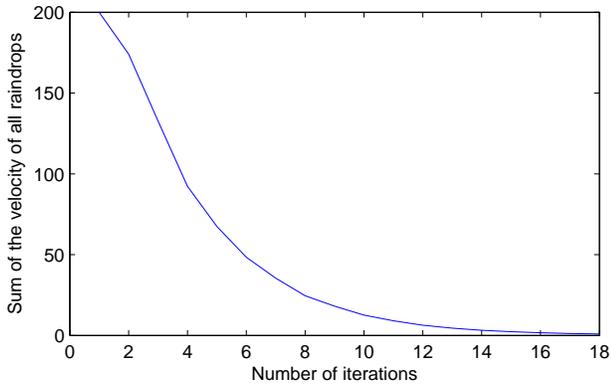}
\caption{Sum of the velocity of all raindrops versus the number of iterations.}
\label{fig_velocity_tends_zero}
\end{figure}

In Fig. \ref{fig_velocity_tends_zero}, 
we illustrate the sum of the velocity of all raindrops 
versus the number of iterations. Notice that with $19$ iterations, 
the velocity of all raindrops are nearly $0$, 
which is the steady state. 
In the steady state, the raindrops are nearly static.

\section{Conclusions}

we apply the random walk model in designing a raindrop algorithm to 
find the global optimal solution of a non-linear programming problem. 
The raindrop algorithm does not require the operation 
of finding the first or second order derivatives of the object function.
Hence this algorithm is a direct method. 
Finally, we apply the raindrop algorithm to solve 
a non-linear optimization problem, 
where the object function is highly irregular (neither convex nor concave). 
And the global optimal solution can be found with limited steps 
of iterations and the success probability is high.

\section*{Acknowledgment}

Thanks Tianping Shuai for the lectures on optimization and algorithm design.

\end{document}